\documentclass{aa}

\usepackage[varg]{txfonts}

\usepackage{graphicx}
\usepackage{natbib}

\begin{document}

\title{A possible common explanation for several cosmic microwave background (CMB) anomalies: A strong impact of nearby galaxies on observed large-scale CMB  fluctuations}

   \author{Frode K. Hansen \inst{1,2}, Ezequiel F. Boero\inst{2,3}, Heliana E. Luparello\inst{2} \and Diego Garcia Lambas\inst{2,3}}

\institute{Institute of Theoretical Astrophysics, University of Oslo, PO Box 1029 Blindern, 0315 Oslo, Norway, \email{frodekh@astro.uio.no}
\and
Instituto de Astronomía Teórica y Experimental, CONICET-UNC, Córdoba, Argentina
\and
Observatorio Astronómico de Córdoba, UNC, Córdoba, Argentina}

%context, aims, methods, results, conclusions

 \abstract
     {A new and hitherto unknown cosmic microwave background (CMB) foreground has recently been detected. A systematic decrease in CMB temperatures around nearby large spiral galaxies points to an unknown interaction with CMB photons in a sphere up to several projected megaparsecs around these galaxies.}
     {We investigate to what extent this foreground may impact the CMB fluctuation map and create the so-called CMB anomalies.}
     {Using the observed temperature decrements around the galaxies, and making some general assumptions about the unknown interaction, we propose a common radial temperature profile. By assigning this profile to nearby galaxies in the redshift range $z=[0.004,0.02],$ we created a foreground map model.}
     {We find a remarkable resemblance between this temperature model map, based on nearby galaxies, and the \textit{Planck} CMB map. Compared to 1000 simulated maps, we find that none of them show such a strong correlation with the foreground map over both large and small angular scales. In particular, the quadrupole, octopole, and $\ell=4$ and $\ell=5$ modes correlate with the foreground map to a high significance. Furthermore, one of the most prominent temperature decrements in the foreground map coincides with the position of the CMB cold spot.}
     {The largest scales of the CMB, and thereby the cosmological parameters, may change significantly after this foreground component  is properly corrected. However, a reliable corrected CMB map can only be derived when suitable physical mechanisms are proposed and tested.}

\authorrunning{Hansen, F. K. et al.} 
\titlerunning{Cosmic microwave background anomalies explained}
 

\maketitle

\section{Introduction}

In \citet[hereafter L2023]{luparello} a systematic decrease in the cosmic microwave background (CMB) temperature was shown around nearby large ($>8.5$\;kpc radius) late spiral galaxies. The authors averaged the radial temperature profile around galaxies with redshift $z<0.015$ from the 2MASS Redshift Survey (2MRS) catalogue \citep{2mrs}, finding a smooth decreasing trend that extends several degrees from the galaxy centres. However, the interaction process and its detailed dependence on galactic properties, such as galaxy type, size, redshift, and level of star formation, is unknown.

Since this extragalactic foreground (hereafter, the `L2023 foreground') extends over large angular scales, we assess whether or not it can provide an explanation of the so-called statistical anomalies in the CMB. Although it is at a low significance level ($2-3\sigma$), the high number of such statistical outliers in WMAP and \textit{Planck} analyses \citep[see][for a recent review]{anomalyrev} warrants further investigation.

The CMB anomalies can be roughly summarised as follows: First, the power spectrum for the lowest multipoles (largest scales) are lower than expected in the best-fit cosmological model, and the multipole modes $\ell<10$ have unexpected features and correlations (see for example \citealt{lowmultipoles}). In particular, the quadrupole and octopole appear to be aligned and similarly dominated by their respective high-$m$ components \citep{alignment}.  Secondly, local estimates of the angular power spectrum, $C_\ell$, indicate a dipolar distribution of power on the sky (\citealt{iands2018} and references therein) with considerably more fluctuation power in the hemisphere centred on $l=246^\circ, b = -2^\circ$ as compared to the opposite hemisphere. This is also reflected in the variance asymmetry \citep{iands2018} since the power spectrum and variance are strongly related and similarly in a dipolar asymmetry in the cosmological parameters (\citealt{paramdipole}, but see also \citealt{paramhorizons}, who point to three separate `horizons' with different cosmological parameters). In \citet{minkowski1}, \citet{minkowski2}, \citet{topology1}, and \citet{topology2}, the topology of the CMB fluctuation field was shown to be anomalous, and the number of cold spots in particular  was found to be anomalously high. Finally, \cite{vielva} showed that the wavelet coefficients for angular scales of about $\sim10^\circ$ on the sky have an excess kurtosis related to a cold spot in the southern galactic hemisphere. Suggested explanations for this anomaly include the presence of a distant large void observed in this region, which could create a cold spot via the integrated Sachs-Wolf effect \cite[see][for a recent update]{void}. However, this structure is not sufficient to explain the observations where imprints on the lensing signal preclude very large under-densities \citep{lensColdSpot}

We modelled the L2023 foreground using some general assumptions about the temperature decrement profile around individual galaxies and fitted the resulting stacked mean profile to the observations. In the resulting foreground temperature map, we looked for features that could explain the aforementioned anomalies. We used publicly available data from the \textit{Planck} satellite experiment\footnote{\texttt{https://www.cosmos.esa.int/web/planck}}, in particular the SMICA foreground-cleaned map \citep{compsep2018} and the corresponding simulations, while the foreground was modelled using galaxies from the 2MASS catalogue, specifically from 2MRS\footnote{\texttt{http://tdc-www.harvard.edu/2mrs/}}  \citep{2mrs}.
 
\section{Modelling the extragalactic foreground}

A detailed modelling of the L2023 foreground is difficult since the nature of the CMB photon interaction with material associated with galaxies is unknown. Nevertheless, we attempted to draw some very general conclusions from the mean observed profiles of different galaxy subsamples. We constructed a simplified foreground model that fits the observed mean temperature profiles around nearby galaxies to study the L2023 foreground impact on the CMB anomalies.

\subsection{Profiles of individual galaxies}

\begin{figure}[htbp]
  \centering
  \includegraphics[scale=0.4]{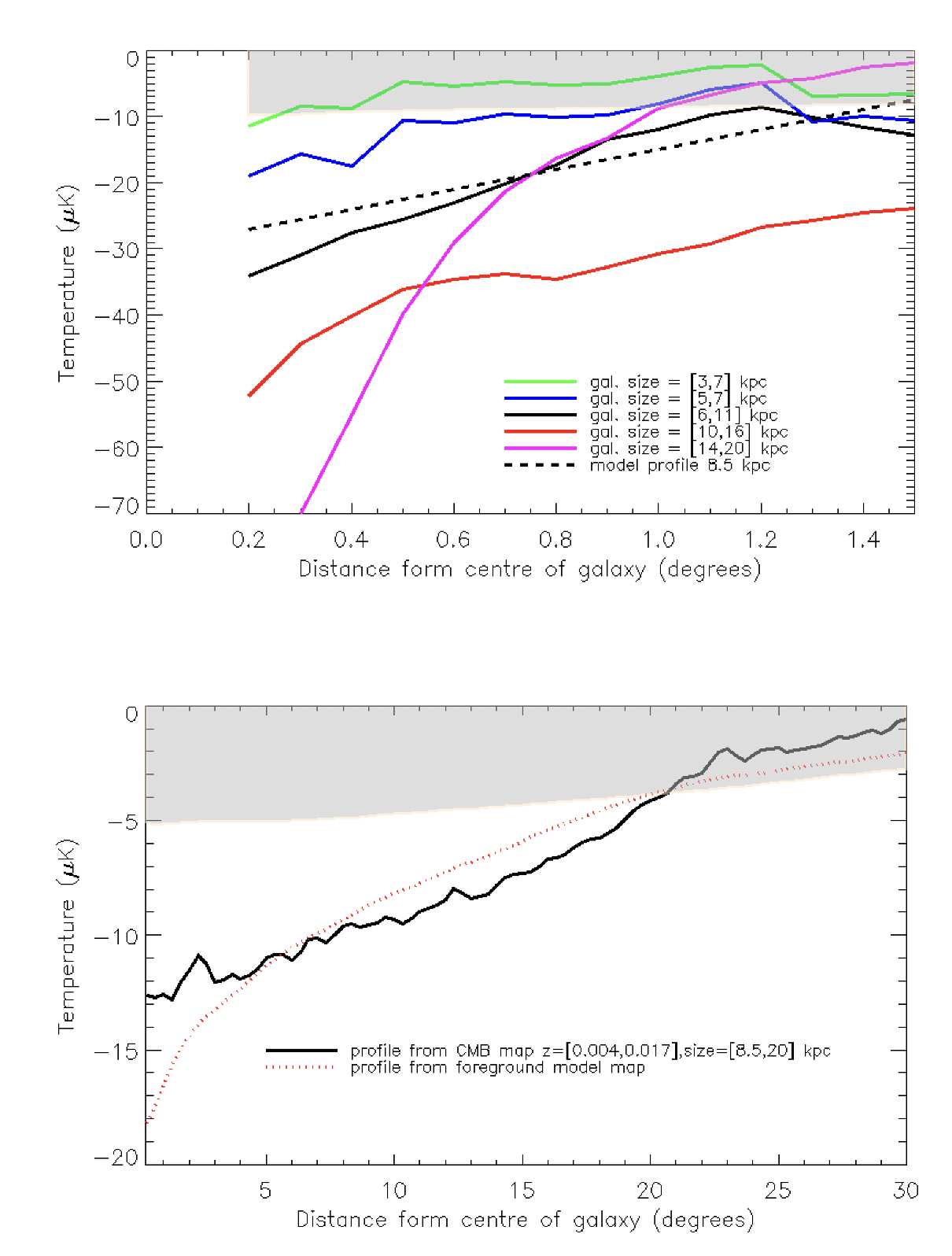}
     \caption{Temperature profiles around 2MRS galaxies. Upper plot: Examples of mean profiles for a subset of isolated late-type spiral galaxies with different size ranges at $z\sim0.01$. The numbers of galaxies in these samples are $\sim 10$ and $\sim 50$ for the smallest and largest sample sizes, respectively. The dashed line shows the model profile that we assign to galaxies of size $8.5$\,kpc. The grey band shows the $1\sigma$ spread of the profiles for simulated CMB maps taken at the position of the galaxies with sizes $[6,11]$\; kpc. Lower plot: Mean profile taken over spiral galaxies within the redshift range $z=[0.004,0.017]$ and the size range $[8.5, 20]$\; kpc for both curves.\ Shown are the observed data (solid line) and the foreground model created with galaxies in the redshift range $[0.004, 0.02]$ (dotted red line). The grey band shows the $1\sigma$ spread of the profile for simulated CMB maps at the same position as the galaxies. }
     \label{fig:profiles}
\end{figure}

In order to create a foreground temperature map, we needed to assign a temperature profile to each galaxy. As discussed in L2023, the signal is mainly associated with late-type spiral galaxies (Sb, Sc, and Sd). A simple linear temperature profile with a given depth and radius was assumed for galaxies of these morphologies, which we obtained using the observed mean temperature profile associated with isolated galaxies. As a measure of galaxy isolation, we used the distance to the fifth closest galaxy, as in L2023. For galaxies smaller than $8.5$\,kpc, we do not find significant temperature decrements for galaxies with distances to the fifth closest galaxy larger than 3 degrees. For the smaller galaxies, we therefore define as `isolated'  those with a distance of between 2.5 and 3 degrees to the fifth closest galaxy. For larger galaxies, we find that we can define the galaxies as isolated if the distance to the fifth closest galaxy is between 3 and 3.5 degrees. More isolated galaxies do not appear to create temperature decrements in the CMB.

We adopted $z=0.01$ and a galaxy size threshold of $8.5$\;kpc as a reference sample to obtain a standard profile that can be scaled to other redshifts, sizes, and environments. In the upper part of Fig. \ref{fig:profiles}, we show profiles for several subsets of isolated galaxies centred on redshift $z=0.01$ for different sizes. For each galaxy size range, the redshift range was adjusted to obtain a clean mean profile within 1-2 degrees of the centre without significant contamination from neighbouring galaxies. Since the \textit{Planck} point source mask removes a disc of radius 0.2 degrees around point sources due to possible contamination, the profile in Fig.\ref{fig:profiles} is restricted to distances $>0.2$ degrees. We notice that by changing the redshift and size ranges, the resulting profile may vary significantly. The example profiles shown correspond to the mean of between $\sim 10$ galaxies (the largest galaxies) and $\sim 50$ galaxies (the smallest galaxies).  We show the adopted reference profile with depth $-30\mathrm{\mu K}$ in the centre and a 2 degree radius. The actual depth and radius uncertainty can be as large as 50\%, although tests on the results show similar outcomes even with such large changes.

\subsection{Creating a foreground model map}

Based on our reference profile, we created a synthetic foreground map, assigning a temperature profile to late spirals in the redshift range $[z=0.004, 0.02]$. Consistent with the upper plot of Fig. \ref{fig:profiles} and L2023, we observe that large spirals have considerably larger temperature decrements than small galaxies. The number of galaxies is too small to accurately quantify the depth profile dependence on galaxy size. Here we adopted a quadratic dependence of the profile depth on galaxy size, making galaxies considerably smaller than $8.5$\;kpc have an almost negligible temperature decrement, consistent with L2023. A quadratic dependence also makes the few largest galaxies ($>20$\;kpc) very dominant. Since we do not know the exact foreground properties, uncertainties in the profile from these few galaxies could contribute to a significant uncertainty in the final foreground map. For this reason, we excluded these extremely large galaxies.

We notice that galaxies in dense environments in general do not have a mean profile depth lower than about $-20 \mu\textrm{K,}$ whereas isolated galaxies normally have even larger depths (see the black line in the lower part of Fig. \ref{fig:profiles} for mean profile over galaxies in all environments). If the mean profile of these galaxies were simply a superposition of the individual profiles, we would expect mean profile depths well below $-20 \mu\textrm{K}$. We suggest that galaxy interaction may be an efficient mechanism by which the unknown material associated with the temperature decrement may be spread over greater distances from the main galaxy, making the depth of the individual galaxies shallower and the radial extension larger in dense environments. This is modelled as a power law dependence of the distance to the fifth closest galaxy for both profile depth and radius. The power law indexes are free parameters obtained by minimising the $\chi^2$ difference between the observed and the modelled mean profile. We find the best-fit power law indexes to be 3.5 for depth and 2.5 for radius.

Since the nearest galaxies with $z<0.004$ have large relative distance uncertainties due to peculiar velocities and also very large angular extensions, they are excluded from our analysis. Introducing variations in the model parameters, profile depth, profile radius, power law indexes, and redshift within the ranges that give suitable fits to the observed mean profiles, our main results are reproduced. We stress the fact that even with a very simple model with a fixed profile to all large spiral galaxies with no other size or density dependence, the qualitative results of our work remain mainly unchanged. Similarly, galaxies with redshifts $z>0.02$ have small angular extensions and contribute minimally to the mean profile.

\begin{figure}[htbp]
  \centering
  \includegraphics[scale=0.4]{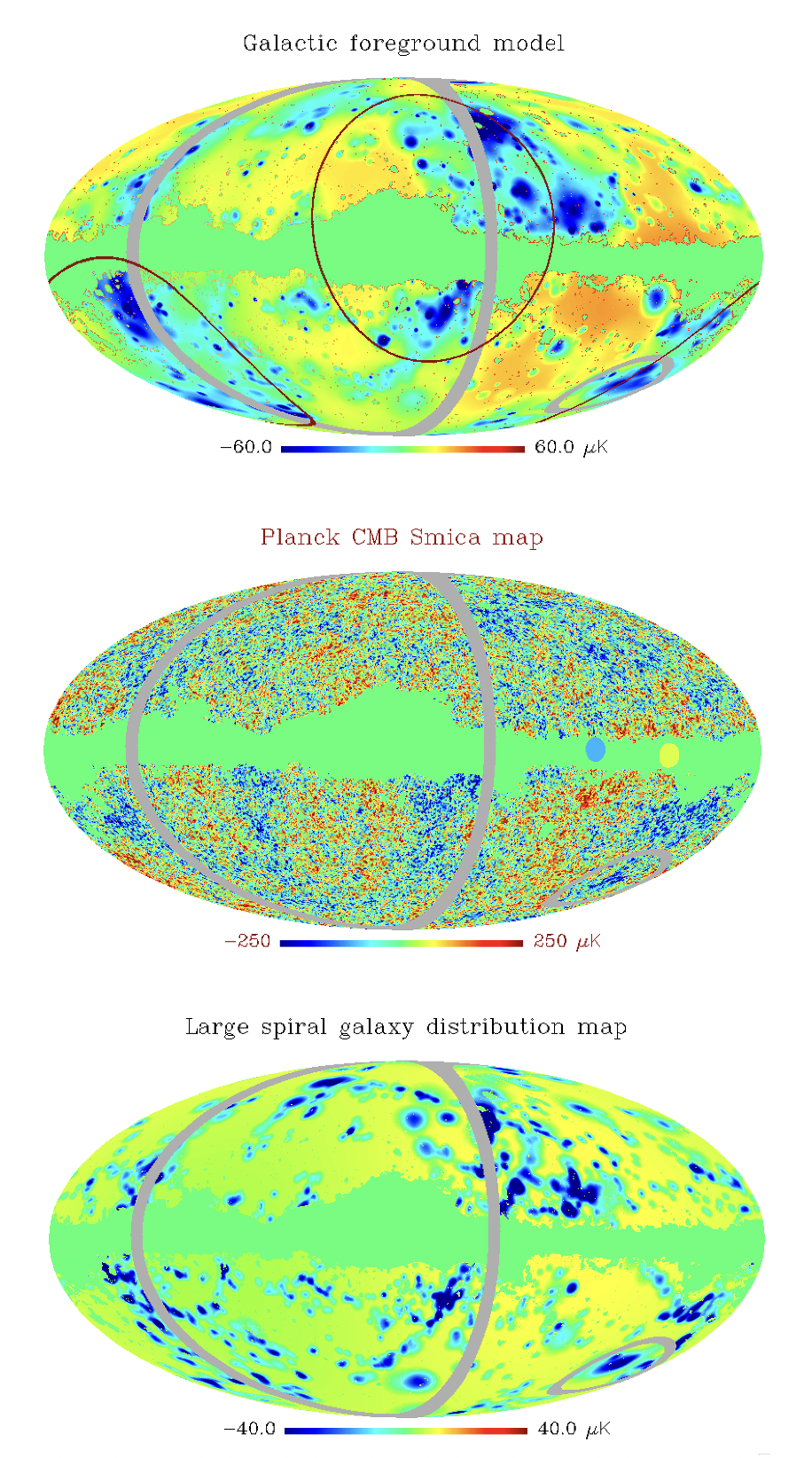}
     \caption{CMB and modelled foreground maps. Top: Foreground model map in $\mu$K. This map was generated by assigning a temperature profile to nearby galaxies ($z<0.02$). The decrement profile was found by making some general assumptions about the L2023 foreground and fitting the mean profile to observations. The two red circles show the parameter `horizons' H1 and H3 from \cite{paramhorizons}. Middle: \textit{Planck} SMICA CMB map. Bottom: Simplified foreground model in which all spiral galaxies $>8.5$\;kpc have been assigned the same profile independently of their size and environment. This is therefore also a density map of nearby spiral galaxies on the sky. In all maps, the grey circle in the lower-right corner shows the position of the cold spot. The maps are divided into two hemispheres, the hemispheres of maximum and minimum power in the multipoles range $\ell=2-220$ obtained from the CMB. In the top CMB map, the blue disc indicates the hemisphere with more power in the foreground map, whereas the yellow disc indicates the centre of the hemisphere with more power in the CMB. }
     \label{fig:maps}
\end{figure}

Our final foreground map is shown in Fig. \ref{fig:maps} (top panel) and the corresponding profile in the lower panel of Fig. \ref{fig:profiles}. The monopole and dipole of the model map have been extracted, consistent with the procedure applied to CMB data, leaving some areas with positive temperatures. The latter plot shows the mean profile of all spiral galaxies in the redshift range $z=[0.004, 0.017]$ with sizes in the range $[8.5,20]$\;kpc, which are the ranges with the strongest foreground signal. We limited the profile redshift to $z=0.017$ as the contributions from higher redshifts are small due to the smaller maximum radius of the profile. We see that the model profile is not a perfect fit to the observations, which highlights the shortcomings of our simple assumptions about the foreground properties. Without a better understanding of the mechanisms that induce the L2023 foreground, the foreground map will have large uncertainties. By means of a simple model, our aim here is to assess if this foreground is associated with most of the observed anomalies. We assume the foreground to be frequency independent, although this will be fully explored in a forthcoming paper.

\section{Assessing anomalies and low-$\ell$ multipoles}

Figure \ref{fig:maps} shows the model foreground map (top panel) and the SMICA \textit{Planck} CMB map (middle panel). Even without a detailed analysis of the anomalies, we already see relevant correlations by comparing the maps. Several of the cold areas in the foreground maps are also cold in the \textit{Planck} CMB map. A grey ring in the lower-right part of the maps points to the non-Gaussian cold spot. We can see that the foreground is expected to cause a considerable temperature decrement at the position of the cold spot. This is one of the dominant temperature decrements in the foreground map and shows that our simple model could already be sufficient to explain a large part of the cold spot anomaly. We will study the cold spot and its contributing galaxies in detail in a separate paper. We also note that the foreground model will increase the number of cold spots in the CMB map. This could explain the results of \citet{minkowski1} and \citet{minkowski2}, who showed that the genus (also known as the Euler characteristic), the number of hot spots minus the number of cold spots, is lower than expected. The low genus suggests that there is an anomalously high number of cold spots relative to the number of hot spots, particularly for scales of 3-4 degrees, which agrees well with our model for nearby galaxies (with an extension of about 2 degrees around galaxies at z=0.01). As the temperature amplitude of our foreground model is of the same order of magnitude as the CMB fluctuations, the large number of disc-like foregrounds with a negative temperature around all the galaxies could significantly increase the number of cold spots. A similar anomaly was found in \cite{topology1} and \cite{topology2} using topological tests.

\begin{figure}[htbp]
\centering
   \includegraphics[scale=0.3]{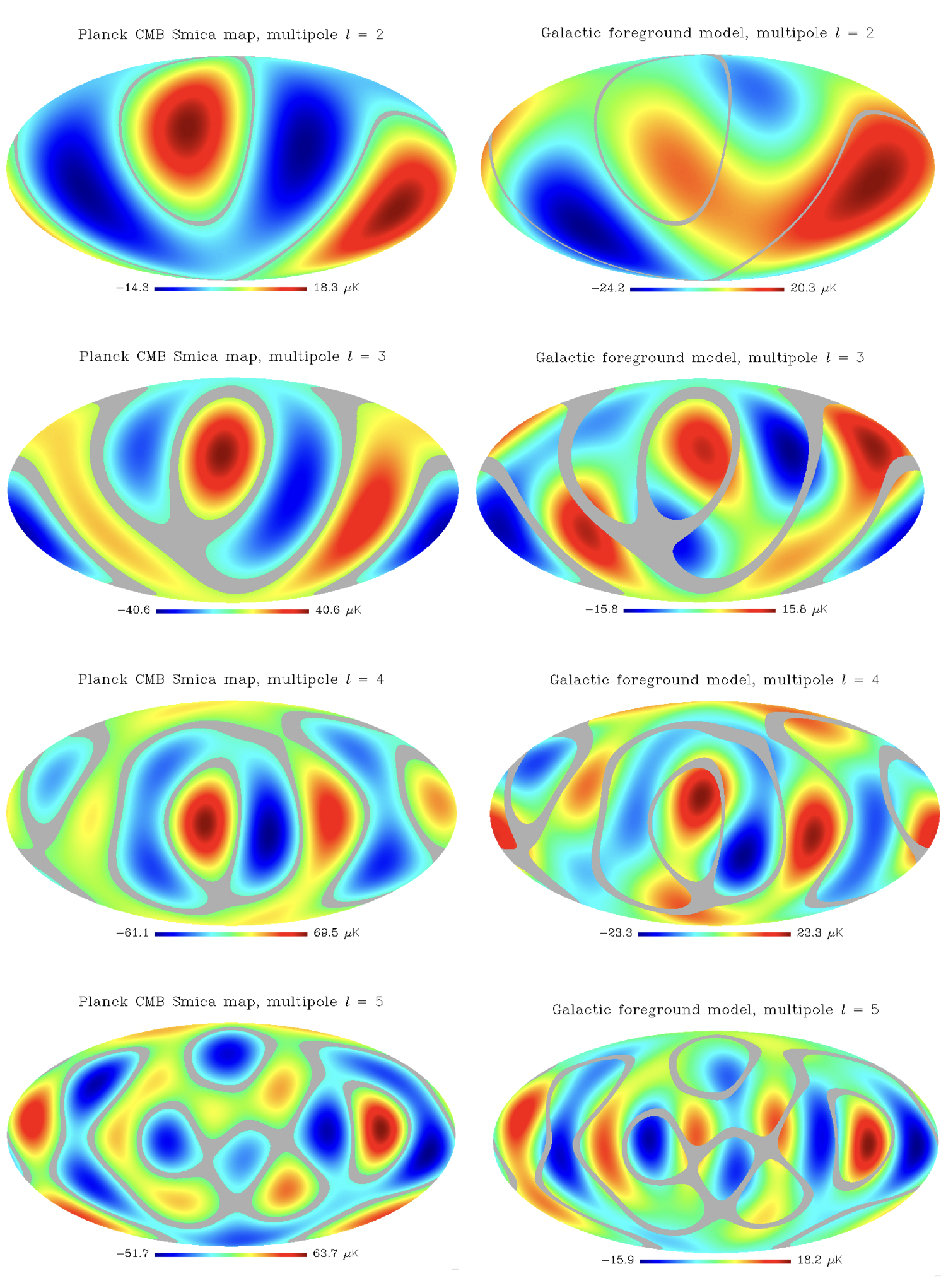}
     \caption{Extraction of the first multipole modes from the maps in Fig. \ref{fig:maps}. The multipoles are estimated outside the \textit{Planck} common mask. Left column: Multipoles of the \textit{Planck} SMICA CMB map.  Right column: Multipoles of the foreground model map. From top to bottom:  $\ell=2$ (quadrupole), $\ell=3$ (octopole), $\ell=4,$ and $\ell=5$ modes. The grey lines serve to aid a by-eye comparison of the position of cold and hot spots between the maps based on the pixels with temperatures close to zero in the CMB map.}
     \label{fig:multipoles}
\end{figure}

Figure \ref{fig:multipoles} shows the quadrupole, octopole, and $\ell=4$ and $\ell=5$ modes of the \textit{Planck} CMB map, in addition to the foreground map. The modes are estimated outside of the \textit{Planck} common galactic mask. The pixels with values close to zero in the \textit{Planck} map are shown in grey to allow for the comparison of the position of hot and cold spots between the CMB and foreground maps. A clear similarity is seen in all four modes, as is the planarity of the octopole (high $m$ domination). Using a standard correlation test, we find that only 0.2\% of simulated maps have a similarly large correlation between the large-scale modes of the foreground maps and the corresponding modes of the CMB.

\begin{figure}[htbp]
\centering
   \includegraphics[scale=0.3]{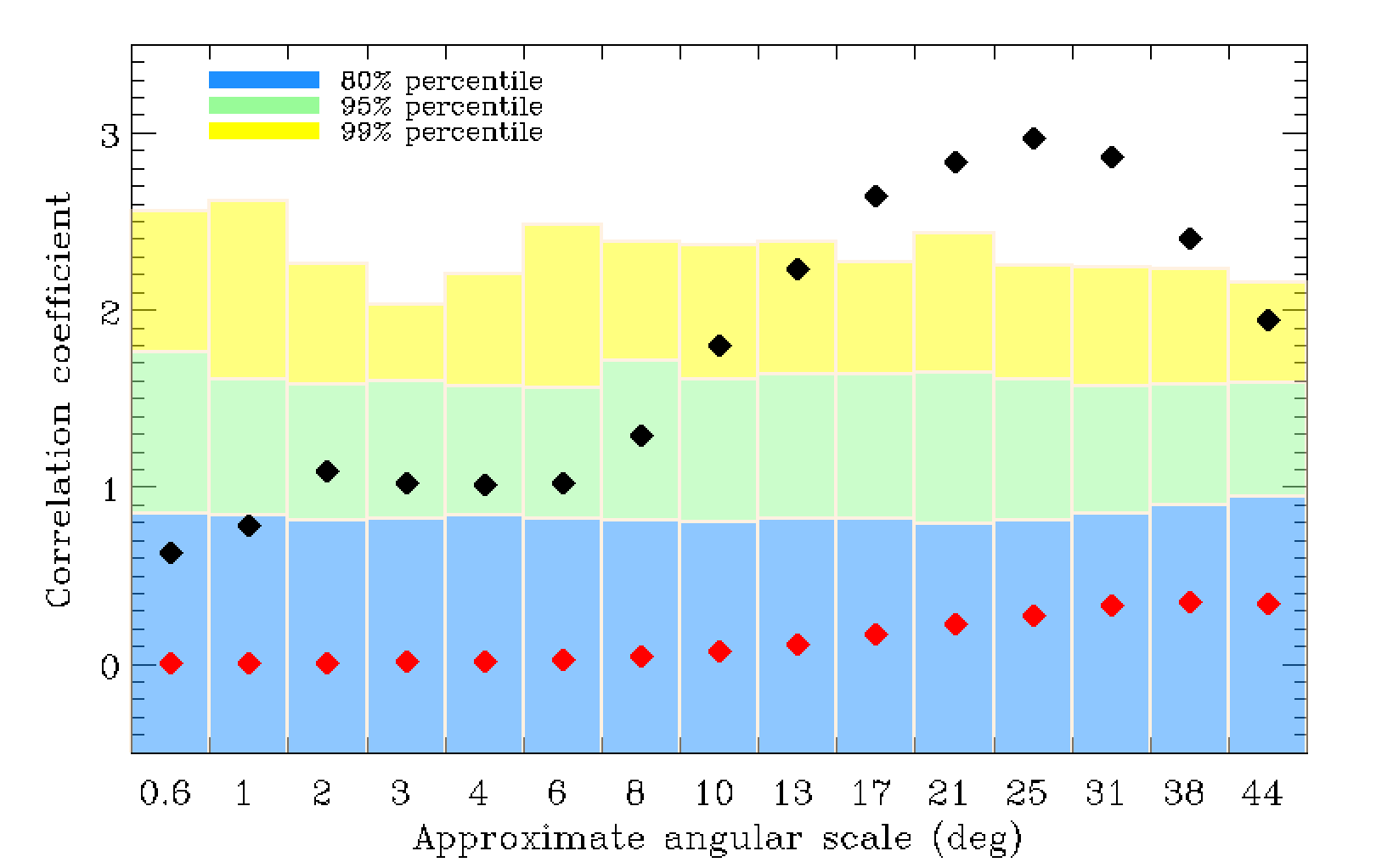}
     \caption{Correlation coefficients between wavelet coefficients for maps of the CMB and the foreground model at the given physical angular scales (corresponding to 2.5 times the spherical Mexican hat wavelet scale). The black points show the correlation coefficients normalised by the standard deviation of the given wavelet scale. The percentiles show the distribution of normalised correlation coefficients of 1000 simulated CMB maps and the foreground model. The red points show the non-normalised correlation coefficients.}
     \label{fig:wavcor}
\end{figure}

In order to test whether this correlation continues to smaller scales, we performed a test of correlations between spherical Mexican hat wavelet coefficients at various scales up to $\ell=1000$ (using the same wavelets and scales as in \citealt{vielva}). We find considerable correlations, as seen in Fig. \ref{fig:wavcor}. In this figure we show the correlation coefficients (red points) as well as the normalised correlation coefficients (black points, divided by the standard deviation of correlation coefficients from 1000 simulations for the given scale). The percentiles show the distribution of normalised correlation coefficients for simulated maps. The larger wavelet scales are correlated with the $\ell=2-5$ multipoles, but scales smaller than $17^\circ$ are less than $20\%$ correlated with the $\ell<6$ modes but still show a significant level of correlation. Only 2\% of the simulated maps have similarly high correlation coefficients for these smaller scales. Also, no simulated map shows a similarly large correlation with the foreground map for both large and small angular scales.

As can be seen in Fig. \ref{fig:maps}, all maps are divided into two hemispheres, the hemispheres with more and less power for the \textit{Planck} CMB map in the range $\ell=2-220$ from \citet{iands2018}. In the foreground model in Fig. \ref{fig:maps}, it  is clear that one hemisphere has considerably more extragalactic structure (and thereby foreground contamination) than the other. Taking the strong correlations between the model foreground map and the \textit{Planck} CMB map into account, we conclude that the foreground signal from all the galaxies in the hemisphere with more structure causes more fluctuations in the corresponding CMB map and is a possible explanation for the hemispherical asymmetry as well as the dipolar asymmetry in the cosmological parameters \citep{paramdipole}. In the top panel of Fig. \ref{fig:maps} we show the horizons H1 and H3 from \citet{paramhorizons}, which are patches where the cosmological parameters exhibit particularly large differences. The third horizon, H2, corresponds relatively well to the direction of hemispherical asymmetry. We can clearly see that there seems to be considerably more foreground contamination (galaxies) inside the three horizons. In fact, the two most extended blue (contaminated) areas of the foreground map coincide to a large degree with the horizons H1 and H3, suggesting that the L2023 foreground could also be responsible for the parameter horizons.

\section{Conclusions}

We have modelled the L2023 foreground \citep{luparello}, creating a simplified map of predicted foreground contamination (see Fig. \ref{fig:maps}). Since we do not understand the interaction process and its properties, the foreground map is only meant as an indication of the expected changes in the CMB temperature fluctuations.

We find that, due to nearby galaxies, the largest-scale fluctuations of the \textit{Planck} CMB map have a remarkable resemblance to those of the foreground map. In particular, the shape of the first multipoles, including the quadrupole and the octopole, shows a strong correlation between the observed CMB and the predicted foreground map (see Fig. \ref{fig:multipoles}). It seems possible that the anomalies associated with the lowest multipoles of the CMB map are to a high degree caused by interactions of CMB photons with extragalactic foregrounds. Using wavelet decomposition, these correlations are seen to continue to smaller angular scales (see Fig. \ref{fig:wavcor}). Taking into account these correlations and the fact that far more nearby galaxies are observed in the hemisphere of maximum power asymmetry in the CMB than in the opposite hemisphere, it seems likely that the unknown foreground component may also be the main cause of the power asymmetry anomaly (and therefore the variance asymmetry) as well as the parameter dipole asymmetry. We have further seen that one of the most dominant cold areas in the foreground map is found at the position of the anomalous CMB cold spot.

We argue that local foregrounds may add to the integrated Sachs-Wolf effect from the Eridanus supervoid (Kovacs et al. 2021) in the same region to produce this compact low temperature area. Given its peculiarity, we will explore it further in a forthcoming publication. Finally, we have seen that the large number of additional cold spots introduced by the foreground around nearby galaxies is a possible explanation for why the Euler characteristic and topological anomalies show an excess number of cold spots with respect to hot spots in the \textit{Planck} and WMAP data. Thus, in addition to large-scale anomalies, the type of foreground explored here may also explain the presence of this excess of compact cold regions without requiring extra assumptions about the CMB. Moreover, testing the topology of the CMB map in the areas where more foregrounds are expected and comparing it to parts of the sky with less expected contamination from nearby galaxies could constitute an important test of the L2023 foreground. Only about 200 of the 2700 galaxies in our model map are masked by 0.2 degree discs by \textit{Planck}; the remaining are unmasked. Masking larger areas around all the nearby galaxies (the cold blue areas of the map in the upper panel of Figure \ref{fig:maps}) in our model and analysing the topology of the remaining area of the sky could reveal whether the extragalactic foreground has a large impact on the topology anomaly.

Based on a simple modelling of an unknown foreground associated with nearby galaxies, we have shown that the largest scales of the CMB seem strongly contaminated. This could pose a significant problem for the interpretation of the CMB power spectrum, since the lowest multipoles will become even smaller after correcting for the L2023 foreground component. It is difficult to see how a very low large-scale power spectrum can be consistent with the standard $\Lambda$ cold dark matter model, but there have been suggestions based on a small causal scale, a loop quantum cosmology, and an ellipsoidal universe (see \citealt{lackcorrelations1, lackcorrelations2, lackcorrelations3} and references therein). We have furthermore shown that the correlations extend to smaller scales and may significantly alter the power spectrum at these scales as well.

We have made general assumptions about the dependence of the foreground temperature profile on galaxy size and environment. We have also made assumptions about the depth, radius, and shape of the reference profile. These assumptions are based on the observed profiles of a few isolated galaxies with limited statistics. The lower part of Fig. \ref{fig:maps} shows a foreground map created by relaxing assumptions about size and environmental dependences. In this map, all spiral galaxies $>8.5$\;kpc have been assigned the same temperature profile, and this is therefore also a map of spiral galaxy density. We see that even a map of the distribution of nearby spiral galaxies is qualitatively similar to the foreground model and also has the same correspondence with the CMB map. For this reason, the qualitative results of this Letter do not strongly depend on the details of the assumptions regarding the extragalactic foreground properties. We also acknowledge the fact that in many cases  there is not an exact coincidence between the galaxy positions and the CMB fluctuations. This is to some degree expected as the hot fluctuations of the CMB in some cases erase the cold imprints of the foregrounds. We further notice that, given the proximity of the foreground tracer galaxies, their peculiar motion in the plane of the sky can account for several degrees during the last billion years. Thus, expected differences in the dynamics of the foreground associated with stripped material and their parent galaxies would also be reflected in the aforementioned lack of exact coincidence.

Understanding the physical mechanisms associated with the L2023 foreground is needed in order to correct the observed CMB maps 
and therefore address its impact on the cosmological parameters. However, given our ignorance of these foreground properties, possible changes in these parameters are currently speculative. We also notice that the limited number of nearby galaxies and the presence of large intrinsic
CMB fluctuations make its modelling difficult. In an upcoming paper, we will study foreground properties in more detail.

\begin{acknowledgements}
We thank the anonymous referee for suggestions that have significantly improved the manuscript.
Results in this paper are based on observations obtained with \textit{Planck}
(http://www.esa.int/Planck), an ESA science mission with instruments
and contributions directly funded by ESA Member States, NASA, and
Canada. The simulations were performed on resources provided by 
UNINETT Sigma2 - the National Infrastructure for High Performance Computing and 
Data Storage in Norway". Some of the results in this paper have been derived using the HEALPix package \citep{healpix}
\end{acknowledgements}

\end{document}